\documentclass[12pt]{article}

\usepackage{color} 
\usepackage[margins]{trackchanges}

\usepackage{amssymb,epsfig}
\usepackage[round]{natbib}
\usepackage[justification=centering]{caption}
\newcommand{\bea}{\begin{eqnarray}}
\newcommand{\eea}{\end{eqnarray}}

\newcommand{\cip}{\perp\!\!\!\perp}

\usepackage[hyphens]{url}

\newtheorem{proposition}{Proposition}

\begin{document}

\author{Vahe Avagyan\\[1ex] 
\textit{Department of Applied Mathematics, Computer Science and Statistics} \textit{\ }%
\\
\textit{Ghent University, Belgium}\\[1ex] 
and Stijn Vansteelandt\\[1ex] 
\textit{Department of Applied Mathematics, Computer Science and Statistics} \textit{\ }%
\\
\textit{Ghent University, Belgium}\\
\textit{and Department of Medical Statistics} \textit{\ }%
\\
\textit{London School of Hygiene and Tropical Medicine, U.K.}}
\title{Honest data-adaptive inference \\ for the average treatment effect using \\ penalised bias-reduced double-robust estimation}
\date{}

\maketitle

\bigskip \setlength{\parindent}{0.3in} \setlength{\baselineskip}{24pt}
\begin{abstract}
The presence of confounding by high-dimensional variables complicates estimation of the average effect of a point treatment. On the one hand, it necessitates the use of variable selection strategies or more general data-adaptive high-dimensional statistical methods. On the other hand, the use of such techniques tends to result in biased estimators with a non-standard asymptotic behaviour. Double-robust estimators are vital for offering a resolution because they possess a so-called small bias property. This means that their bias vanishes faster than the bias in the nuisance parameter estimators when the relevant smoothing parameter goes to zero, provided that certain sparsity assumptions hold. This property has been exploited to achieve valid (uniform) inference of the average causal effect when data-adaptive estimators of the propensity score and conditional outcome mean both converge to their respective truths at sufficiently fast rate \citep[e.g., ][]{farrell2015robust, belloni2016post}. In this article, we extend this work in order to retain valid (uniform) inference when one of these estimators does not converge to the truth, regardless of which. This is done by generalising prior work by \cite{vermeulen2015bias} to incorporate regularisation. The proposed penalised bias-reduced double-robust estimation strategy exhibits promising performance in extensive simulation studies and a data analysis, relative to competing proposals.
\end{abstract}


\section{Introduction}
\label{Section:1}

The effects of treatments, policies or interventions are commonly characterised in terms of contrasts between the mean of counterfactual outcomes corresponding to different treatment or exposure levels. For instance, for a dichotomous treatment $A$ (coded 0 for no treatment and 1 for treatment), the average treatment effect (ATE) is defined as $E\left\{Y(1)\right\}-E\left\{Y(0)\right\}$, where $Y(a)$ denotes the counterfactual outcome of a random individual if that individual were exposed to treatment $a=0,1$. Estimation of such effect from observational data generally requires adjustment for a set of covariates that are sufficient to adjust for confounding of the effect of treatment on outcome. 
This is a difficult task when the number of covariates is large or when one or multiple continuous  covariates can have non-linear effects on exposure or outcome. It is therefore common to start from flexible models and adopt variable selection or 
more general regularisation techniques to handle the high dimensionality of the models. Such data-adaptive techniques are especially crucial when the number of variables $p$ is large relative to the number of observations $n$. 

The use of data-adaptive techniques requires consideration in itself, however. Regularisation techniques tend to return biased estimators (e.g. for the dependence of treatment or outcome on covariates). Estimators of the ATE based on these, may inherit this bias. Nuisance parameter estimators obtained via regularisation techniques also typically have a non-normal asymptotic distribution \citep{knight2000asymptotics, leeb2005model}. This may render the distribution of ATE estimators based on these rather complicated. Both these concerns make asymptotically unbiased estimators for the ATE with accompanying uniformly valid confidence intervals difficult to attain, especially in settings where the models' complexity increases with sample size. This forms one of the major Achilles heels of routine data analyses, since uniform validity is essential in order to trust their finite-sample performance.

So-called double-robust (DR) estimators of the ATE (\citeauthor{robins2001comments}, \citeyear{robins2001comments}; see \citeauthor{rotnitzky2014double}, \citeyear{rotnitzky2014double} for a review) are not susceptible to the above problems, under certain conditions that we will specify next. 
DR estimators of the ATE make use of two working models: one model $\mathcal{A}$ for the dependence of exposure on covariates, and one model $\mathcal{B}$ for the dependence of outcome on covariates. They have the attractive property of being consistent for the ATE when either one of these working models is correctly specified, but not necessarily both. 
When both nuisance working models  $\mathcal{A}$ and $\mathcal{B}$ are correctly specified and estimated at faster than $n^{-1/4}$ rate (in a sense to be made precise later), then DR estimators of the ATE are orthogonal (w.r.t. the covariance inner product) to the scores for the infinite-dimensional nuisance parameters that index the observed data distribution (i.e., the probability of treatment given covariates, and the outcome distribution given covariates and fixed treatment levels). This in turns implies that estimation (and in particular, regularisation) of these nuisance parameters can be ignored and, hence, that the resulting DR estimator is asymptotically unbiased with standard, easy-to-calculate confidence interval that is uniformly valid \citep{van2014targeted, farrell2015robust, belloni2016post, athey2016efficient}. This surprising result applies to any (sufficiently fast converging) data-adaptive method for estimating nuisance parameters; in particular, it forms the cornerstone of the now popular Targeted Maximum Likelihood method \citep{van2011targeted}.

While promising, a limitation of the above result is that it assumes both nuisance working models  $\mathcal{A}$ and $\mathcal{B}$ to be correctly specified (or more generally, both nuisance parameter estimators to converge to their respective truths). This is unlikely to be satisfied. Current practice is often based on simple parametric working models. Moreover, the data analyst is essentially always forced to constrain the model's flexibility in order to ensure nuisance parameter estimators that are sufficiently fast converging. In view of this, in this article, we will generalise the above results to allow for misspecification of both nuisance working models  $\mathcal{A}$ and $\mathcal{B}$. In particular, we will show that the use of special nuisance parameter estimators will yield a DR estimator 
which is asymptotically unbiased when at least one of the working models is correctly specified, and will moreover yield an accompanying Wald confidence interval that is easy to calculate and uniformly valid for the estimator's probability limit, even when both working models are misspecified. We will achieve this goal by extending the bias-reduced DR estimation principle of \cite{vermeulen2015bias} to incorporate regularisation in a way that is inspired by penalised estimation equations \citep{fu1998penalized}. In particular, we will consider $\ell_1$ or Lasso norm penalisation \citep{tibshirani1996regression, fu2003penalized} in order to prevent slowly converging, and therefore potentially severely biased estimators, which may otherwise result when the working models include many (unimportant) covariates. 

The rest of the article is organised as follows. In Section \ref{Section:3}, we describe our proposed penalised bias-reduced DR estimator and evaluate its asymptotic properties. We explore connections to earlier work on bias-reduced DR estimation in low-dimensional settings in Section \ref{Section:2}. In Section \ref{Section:4}, we numerically evaluate the performance of the proposed estimators in comparison with other DR estimators through extensive simulation studies, as well as with an ad hoc extension based on double-selection \citep{belloni2013honest, belloni2016post}. We illustrate the proposed estimators in an application on the effect of life expectancy on economic growth in 
Section \ref{Section:5} and conclude with suggestions for future work in section \ref{Section:6}. 

\section{Penalised Bias-Reduced Double-Robust Estimation}
\label{Section:3}
\subsection{Background}

Consider a study design which intends to collect i.i.d. data on an outcome $Y_i$, a treatment $A_i$ (coded 0 or 1) and a $p$-dimensional vector of covariates $X_i$ for subjects $i=1,...,n$. Our focus will be on the estimation of the counterfactual mean $\mu_0\equiv E\{Y(1)\}$  under the nonparametric model $\mathcal{M}$ for the observed data $(Y,A,X)$, which is defined by the assumption that $X$ is sufficient to control for confounding of the exposure effect, in the sense that $Y(1)\cip A|X$, and 
the so-called consistency assumption that the conditional laws of $Y$ and $Y(1)$, given $A=1$ and $X$, are identical. 
Throughout, we will also make the positivity assumption that $P(A=1|X)\in [\delta,1-\delta]$ for some $\delta>0$ with probability 1.
Note that $E\{Y(1)\}$ is one component of the ATE; estimation of $E\{Y(0)\}$ proceeds analogously upon changing the treatment coding. 

Unless $X$ is limited to few (e.g. one or two) discrete covariates, some form of dimension reduction 
is typically needed in order to obtain a well-behaved estimator of the marginal treatment effect in small to moderate sample sizes (Robins and Ritov, 1997). 
For instance, in routine practice, it is common to adjust for confounding under a low-dimensional model for the dependence of $X$ on the outcome. In particular, in this article we will proceed under the assumption that the expected outcome in exposed obeys a parametric (working) model $\mathcal{B}$, which postulates that $E(Y|A=1,X)=m(X;\beta^*)$ where $m(X;\beta)$ is a known function, smooth in $\beta$, and $\beta^*$ is unknown, e.g. $m(X;\beta)=\beta_0+\beta_1X+\beta_2X^2$ with $\beta\equiv (\beta_0,\beta_1,\beta_2)'$. Given a consistent estimator $\hat{\beta}$ of $\beta^*$, $\mu_0$ can then be estimated as 
\[\tilde{\mu}=\frac{1}{n}\sum_{i=1}^n m(X;\hat{\beta}).\]

In high-dimensional settings where the number of covariates $p$ is large relative to the sample size $n$ (i.e., $p$ is allowed to grow with $n$), data-adaptive procedures (e.g. stepwise variable selection, Lasso or more general penalisation procedures, among others) cannot usually be avoided for estimating the conditional outcome mean. These procedures typically return biased estimators, as a result of sparsity in the data and the resulting need to regularise. The estimator $\tilde{\mu}$ may inherit this bias \citep{bickel1982adaptive} and, moreover, follow a non-standard asymptotic distribution as a result, making uniformly valid confidence intervals for $\mu_0$ difficult to attain (see Section \ref{subsec:asympt} for detail).
 
DR estimators of $\mu_0$ form an exception \citep{belloni2012sparse, van2014targeted, farrell2015robust}. In particular, 
let $\mathcal{A}$ be a parametric working model $P(A=1|X)=\pi(X;\gamma^*)$ for the probability of being exposed, where 
$\pi(X;\gamma)$ is a known function, smooth in $\gamma$, and $\gamma^*$ is unknown,
e.g. $\pi(X;\gamma)=1/\left\{1+\exp(-\gamma_0-\gamma_1X)\right\}$ with $\gamma\equiv (\gamma_0,\gamma_1)'$. Consider now the estimator 
\[\hat{\mu}=\frac{1}{n}\sum_{i=1}^n U_i(\hat{m},\hat{\pi}),\]
with 
\begin{equation}\label{ex2}
U(m,\pi)\equiv m(X)+\frac{A}{\pi(X)}\left\{Y-m(X)\right\},\end{equation}
where $m(X)\equiv E(Y|A=1,X)$ and $\pi(X)\equiv P(A=1|X)$, and $\hat{m}(X)$ and $\hat{\pi}(X)$ are data-adaptive fits of $m(X)$ under model $\mathcal{B}$ and $\pi(X)$ under model $\mathcal{A}$, respectively. 
This estimator is double-robust in the sense that it converges to $\mu_0$ when either $\hat{m}(X)$ converges to $E(Y|A=1,X)$ or $\hat{\pi}(X)$ converges to $P(A=1|X)$, but not necessarily both. It follows from \cite{farrell2015robust} that $\hat{\mu}$ has the same asymptotic distribution as $n^{-1}\sum_{i=1}^n U_i(m,\pi)$, regardless of the choice of estimators $\hat{m}(X)$ and $\hat{\pi}(X)$, provided that both are consistent and that the product of their sample mean squared errors converges at faster than $n$ to the quarter rate. Uniformly valid, normal confidence intervals for $\mu_0$ are therefore straightforwardly obtained based on a standard error which can be consistently estimated as 1 over $n$ times the sample variance of $U(m,\pi)$, evaluated at $m(X)=\hat{m}(X)$ and $\pi(X)=\hat{\pi}(X)$ \citep{farrell2015robust}. 

Unfortunately, consistent estimation of both $m(X)$ and $\pi(X)$ is unlikely in high-dimensional settings (where $p$ may even grow with $n$). Indeed, the sparsity in the data necessitates one to make simplifying assumptions, such as the parametric model restrictions $\mathcal{A}$ or $\mathcal{B}$, in order to obtain fast enough converging estimators. Such restrictions are unlikely to be entirely correct. In this paper, we therefore aim to obtain uniformly valid standard errors, even under misspecification. We will first explain the procedure, and then demonstrate its asymptotic properties in the next section. 
 
\subsection{Proposal}\label{proposal}

As in \cite{belloni2012sparse} and \cite{farrell2015robust}, we will develop inference for $\mu_0$ under 
parametric working models with high-dimensional covariates (where $p$ may potentially exceed $n$).
Our proposal is then to estimate $\mu_0$ as $\hat{\mu}=\frac{1}{n}\displaystyle\sum_{i=1}^n U_i(\hat{\eta})$ for a nuisance parameter estimator 
$\hat{\eta}=(\hat{\gamma}',\hat{\beta}')'$ obtained
by solving the following penalised estimating equations using the bridge penalty \citep{fu2003penalized}:
\[0=\left[\frac{1}{n}\sum_{i=1}^n \frac{\partial}{\partial \beta} U_i(\hat{\eta}),\frac{1}{n}\sum_{i=1}^n \frac{\partial}{\partial \gamma} U_i(\hat{\eta})\right]+\left[\lambda_{\gamma} \delta |\hat{\gamma}|^{\delta-1}\circ\mbox{\rm sign}(\hat{\gamma}),\lambda_{\beta} \delta |\hat{\beta}|^{\delta-1}\circ\mbox{\rm sign}(\hat{\beta})\right],\]
where $\lambda_{\gamma} > 0$ and $\lambda_{\beta} > 0$ are the associated penalty parameters and $\delta\ge 1$.
Here, for vectors $a\in \mathbb{R}^p$ and $b\in \mathbb{R}^p$, $c=a\circ b \in  \mathbb{R}^p$ refers to the so-called elementwise (or Hadamard) product, where $c=(c_1,...,c_p)$ with $c_i=a_ib_i$ for $i=1,...,p$. Further, $\textnormal{sign}(a)$ for a vector $a\in \mathbb{R}^p$ is defined as a vector of elements $\textnormal{sign}(a_j)$, for $j=1,...,p$
Finally, the terms $\delta |\hat{\gamma}|^{\delta-1}\circ\mbox{\rm sign}(\hat{\gamma})$ and $\delta |{\beta}|^{\delta-1}\circ\mbox{\rm sign}({\beta})$ are the partial derivatives of $||\gamma||_\delta^{\delta}$ and $||\beta||_\delta^{\delta}$ with respect to $\gamma$ and $\beta$, respectively, where the $\ell_\delta$ norm is defined as $||a||_\delta\equiv\left({\displaystyle\sum_{i=1}^p |a_i|^\delta}\right)^{1/\delta}$. 

Throughout, for pedagogic purposes, we will specialise our proposal to working models of the form
\[\pi(X;\gamma)= \mbox{expit}(\gamma' (1,X)),\]
and 
\[m(X;\beta)=\beta' (1,X).\]
In that case, we first solve the set of penalised estimating equations:
\begin{eqnarray}
\label{EstEq1}
0&=&\frac{1}{n}\sum_{i=1}^n \frac{\partial}{\partial \beta} U_i(\hat{\eta})+\lambda_{\gamma} \delta |\hat{\gamma}|^{\delta-1}\circ\mbox{\rm sign}(\hat{\gamma})\nonumber\\
&=&\frac{1}{n}\sum_{i=1}^n \left\{1-\frac{A_i}{\pi({X_i,\hat{\gamma}})}\right\}(1,X_i')'+\lambda_{\gamma} \delta |\hat{\gamma}|^{\delta-1}\circ\mbox{\rm sign}(\hat{\gamma}).
\end{eqnarray}
to estimate $\gamma$. For $\delta\rightarrow 1+$, the penalty term $\delta |\hat{\gamma}|^{\delta-1}\circ\mbox{\rm sign}(\hat{\gamma})$ has $j$th component 
$\textnormal{sign}(\hat{\gamma}_j)$ if $\hat{\gamma}_j\not=0$ and belongs to $[-1,1]$ otherwise (see Section 3 of supplementary materials for more details). In that case, we recommend solving this equation by minimising the function \citep{vermeulen2015bias}:
\begin{eqnarray}
\label{OptimProb1}
\min_{\gamma} \mathcal{F}_1(\gamma)&=&\frac{1}{n}\sum_{i=1}^n\left[A_i\exp(-\gamma'(1,X_i')')+(1-A_i)\gamma'(1,X_i')'\right]+{\lambda_{\gamma}} ||\gamma||_1.
\end{eqnarray}
This results in an estimator $\hat{\gamma}$ of $\gamma$.

We next solve the set of penalised estimating equations:
\begin{eqnarray}
\label{EstEq2}
0&=&\frac{1}{n}\sum_{i=1}^n \frac{\partial}{\partial \gamma} U_i(\hat{\eta})+\lambda_{\beta} \delta |\hat{\beta}|^{\delta-1}\circ\mbox{\rm sign}(\hat{\beta})\nonumber\\
&=&-\frac{1}{n}\sum_{i=1}^n \hat{w}_iA_i\left\{Y_i-m(X_i, \hat{\beta})\right\}(1,X_i)+\lambda_{\beta} \delta |\hat{\beta}|^{\delta-1}\circ\mbox{\rm sign}(\hat{\beta}),
\end{eqnarray}
where 
\[\hat{w}_i\equiv \frac{1-\pi(X_i, \hat{\gamma})}{\pi(X_i, \hat{\gamma})}>0.\]
For $\delta=1$, this is best done by minimising the function:
\begin{eqnarray}
\label{OptimProb2}
\min_{\beta} \mathcal{F}_2(\beta)&=&\frac{1}{2n}\sum_{i=1}^n\left[\hat{w}_iA_i(Y_i-m(X_i,\beta))^2\right]+{\lambda_{\beta}} ||\beta||_1,
\end{eqnarray}
which is possible by standard software for (weighted) $\ell_1$-penalisation. This results in an estimator $\hat{\beta}$ of $\beta$.

The above proposal generalises the bias-reduced DR estimation procedure of \cite{vermeulen2015bias} to incorporate penalisation. In low-dimensional settings with $\lambda_{\gamma}=\lambda_{\beta}=0$, it delivers consistent nuisance parameter estimators under correct model specification. However, it requires nuisance parameters $\beta$ and $\gamma$ of equal dimension, since the gradient ${\partial} U({\eta})/{\partial \beta}$ (for ${\eta}=({\gamma}',{\beta}')'$) carries information about $\gamma$, and vice versa, the gradient ${\partial} U({\eta})/{\partial \gamma}$ carries information about $\beta$ \citep{vermeulen2015bias}. This limitation is essentially resolved by letting $\delta\to 1+$ \citep{fu2003penalized}. This makes the penalty terms correspond to the sub-gradient of the $\ell_1$ or Lasso norm penalty $||\eta||_1$ with respect to $\eta$ \citep{tibshirani1996regression}, thereby guaranteeing both convexity and sparsity, and thus possibly resulting in nuisance parameter estimates with different numbers of non-zero components.
In the next section, we will demonstrate that the above proposal enables uniformly valid inference in high-dimensional settings where either model $\mathcal{A}$ or $\mathcal{B}$ - but not both - is misspecified. 

\subsection{Asymptotic properties}\label{subsec:asympt}

As in \cite{belloni2012sparse} and \cite{farrell2015robust}, we will study convergence under an arbitrary sequence $\{P_n\}$ of observed data laws that obey, at each $n$, the positivity assumption. This implies that the parameters $\eta$ and $\mu_0$, as well as the models $\mathcal{M},\mathcal{A}$ and $\mathcal{B}$ should ideally be indexed by $n$, although we will suppress this notation where it does not raise confusion. Allowing for such dependence on $n$ is quite natural because we are considering settings where the number of covariates, and thus the dimension of $\eta$, may increase with sample size \citep{farrell2015robust}. It is also required in order to demonstrate uniform convergence, as we will argue below. 

We will furthermore consider settings where the working models $\mathcal{A}$ and $\mathcal{B}$ may be misspecified. The population value of the nuisance parameter $\eta$ may thus be ill-defined and we will therefore study (the rate of) convergence of $\hat{\eta}$ to the solution $\eta^*_n\equiv (\gamma_n^{*'},\beta_n^{*'})'$ to the population equation
\[E_{P_n}\left\{\frac{\partial U}{\partial \eta} (\eta)\right\}=0,\]
where we make explicit that the expectation is taken w.r.t. the law $P_n$. It follows from \cite{vermeulen2015bias} that the component $\gamma_n^*$ equals the population value of $\gamma$ indexing model $\mathcal{A}$ (under the law $P_n$)  when that model is correctly specified, and likewise that the component $\beta_n^*$ equals the population value of $\beta$ indexing model $\mathcal{B}$ (under the law $P_n$) when that model is correctly specified. Our main result in Proposition \ref{MainProposition} below now states that $n^{-1/2}\sum_{i=1}^n U_i(\hat{\eta})$ and $n^{-1/2}\sum_{i=1}^n U_i({\eta^*_n})$ are asymptotically equivalent under model $\mathcal{M}$, even under the `worst' sequence of laws $P_n$ and even when the working models $\mathcal{A}$ and $\mathcal{B}$ are misspecified, provided that certain sparsity assumptions hold. Under these assumptions, we thus have that
\begin{eqnarray*}
\sqrt{n}(\hat{\mu}-\mu_0)&=&\frac{1}{\sqrt{n}}\left\{\sum_{i=1}^nU_i(\hat{\eta})-U_i(\eta^*)+U_i(\eta^*)-\mu_0\right\}\\
&=&\frac{1}{\sqrt{n}}\sum_{i=1}^n\left\{U_i(\eta^*)-\mu_0\right\}+\frac{1}{\sqrt{n}}\sum_{i=1}^n\left\{U_i(\hat{\eta})-U_i(\eta^*)\right\}\\
&=&\frac{1}{\sqrt{n}}\sum_{i=1}^n\left\{U_i(\eta^*)-\mu_0\right\}+o_{P_n}(1),
\end{eqnarray*}
where the term $o_{P_n}(1)$ converges to zero in probability under the measure $P_n$.
It follows from this that 
the uncertainty in the estimator $\hat{\eta}$ can be ignored when doing inference about $\mu_0$, and in particular that a uniformly consistent estimator of the standard error of $\hat{\mu}$ can be obtained as $\hat{\sigma}/\sqrt{n}$, with 
\[\hat{\sigma}=\left(\frac{1}{n-1}\sum_{i=1}^n \left\{U_i(\hat{\eta})-\hat{\mu}\right\}^2\right)^{1/2}.\]
It further follows from the above proposition that, when either model $\mathcal{A}$ or model $\mathcal{B}$ is correctly specified so that  $\hat{\mu}$ converges to $\mu_0$, a uniformly valid confidence interval for $\mu_0$ can be obtained as
\[\hat{\mu}\pm 1.96\hat{\sigma}/\sqrt{n}.\]


\begin{proposition}
\label{MainProposition}
Let $\hat{\eta}$ be the estimator of $\eta=(\gamma',\beta')'$ as obtained via the proposed penalised bias-reduced DR method. 
Define the active set of the variables as $S_{\gamma}=\textnormal{supp}(\gamma^*_n)$, $S_{\beta}=\textnormal{supp}(\beta^*_n)$,
where, for any vector $a\in \mathbb{R}^p$, we denote its support as $\textnormal{supp}(a)=\{i\in \{1,...,p\}| a_i\not=0\}$. Let the sparsity index 
$s_\gamma$ equal the cardinality $|S_{\gamma}|$, and likewise $s_\beta=|S_{\beta}|$; note that $s_\gamma$ and $s_\beta$ may depend on $n$.
If $\lambda_{\gamma}=O\left(\sqrt{\frac{\log{p}}{n}}\right)$ and $\lambda_{\beta}=O\left(\sqrt{\frac{\log{p}}{n}}\right)$ and the assumptions in Section 1 of supplementary materials hold, then 
\begin{eqnarray*}
\Bigr|\frac{1}{\sqrt{n}}\sum_{i=1}^n \left(U_i(\eta^*)-U_i(\hat{\eta})\right)\Bigr|
&=&O_{P_n}\left\{(s_{\gamma}+s_{\beta})\frac{\log p}{\sqrt{n}}\right\}.
\end{eqnarray*}
Provided sufficient sparsity in the sense that $(s_{\gamma}+s_{\beta})\log p/\sqrt{n}$ converges to zero with increasing sample size, it follows that 
\[\lim_{n\rightarrow\infty}\sup_{P_n} P_n\left\{\Bigr| n^{-1/2}\sum_{i=1}^n U_i(\hat{\eta})-n^{-1/2}\sum_{i=1}^n U_i({\eta^*})\Bigr|>\epsilon\right\}=0,\]
under model $\mathcal{M}$, even when the working models $\mathcal{A}$ and $\mathcal{B}$ are misspecified. 
\end{proposition} 

Below we give the key part of the proof of Proposition \ref{MainProposition}, which is instructive to understand the logic behind the proposed method. Further details are given in Section 1 of supplementary materials.

\textit{Proof:} The proof of Proposition \ref{MainProposition} follows similar lines as in \cite{ning2017general}. Taylor expansion shows that
\begin{eqnarray*}
\frac{1}{\sqrt{n}}\sum_{i=1}^n U_i(\eta^*_n)
&=&\frac{1}{\sqrt{n}}\sum_{i=1}^n U_i(\hat{\eta})-\frac{1}{n}\sum_{i=1}^n \frac{\partial U_i}{\partial \gamma} (\hat{\eta}) \sqrt{n}(\hat{\gamma}-{\gamma}^*_n)\\
&&-\frac{1}{n}\sum_{i=1}^n \frac{\partial U_i}{\partial \beta} (\hat{\eta}) \sqrt{n}(\hat{\beta}-{\beta}^*_n)+O_{P_n}(\sqrt{n}\|\hat{\eta}-{\eta}^*_n\|^2_2).
\end{eqnarray*}
Let for any vector $a=(a_1,...,a_p)\in \mathbb{R}^p$, $||a||_{\infty}=\max_{i}|a_i|$ denote the $\ell_\infty$ or sup norm. Then from H\"older's inequality we have
\begin{eqnarray*}
\Bigr|\frac{1}{n}\sum_{i=1}^n \frac{\partial U_i}{\partial \gamma} (\hat{\eta}) \sqrt{n}(\hat{\gamma}-{\gamma}^*_n)\Bigr|&\leq &
\Bigr|\Bigr|\frac{1}{n}\sum_{i=1}^n \frac{\partial U_i}{\partial \gamma} (\hat{\eta})\Bigr|\Bigr|_{\infty}\| \sqrt{n}(\hat{\gamma}-{\gamma}_n^*)\|_1\\
&=&\|\lambda_{\beta} \delta |\hat{\beta}|^{\delta-1}\mbox{\rm sign}(\hat{\beta})\|_{\infty}\| \sqrt{n}(\hat{\gamma}-{\gamma}_n^*)\|_1\\
&\le&\lambda_{\beta} \delta \|\sqrt{n}(\hat{\gamma}-{\gamma}_n^*)\|_1,
\end{eqnarray*}
since $||\delta |\hat{\beta}|^{\delta-1}\mbox{\rm sign}(\hat{\beta}) ||_\infty\le 1$ (for $\delta\to 1+)$, and likewise that 
\begin{eqnarray*}
\Bigr|\frac{1}{n}\sum_{i=1}^n \frac{\partial U_i}{\partial \beta} (\hat{\eta}) \sqrt{n}(\hat{\beta}-{\beta}^*_n)\Bigr|&\leq &\lambda_{\gamma} \delta\| \sqrt{n}(\hat{\beta}-{\beta}_n^*)\|_1.
\end{eqnarray*}
Suppose now that 
\begin{eqnarray*}
\lim_{n\rightarrow\infty} P_n\left\{\| \hat{\eta}-{\eta}_n^*\|_2\lesssim c_2(n)\right\}&=&1\\
\lim_{n\rightarrow\infty} P_n\left\{\| \hat{\gamma}-{\gamma}_n^*\|_1\lesssim c_{1\gamma}(n)\right\}&=&1\\
\lim_{n\rightarrow\infty} P_n\left\{\| \hat{\beta}-{\beta}_n^*\|_1\lesssim c_{1\beta}(n)\right\}&=&1,
\end{eqnarray*}
where $c_{1\gamma}(n),c_{1\beta}(n)$ and $c_2(n)$ converge to zero as $n\rightarrow\infty$; here, for positive sequences $a_n$ and $b_n$, we use the notation $a_n\lesssim b_n$ to denote $a_n\leq Cb_n$ for some constant $C>0$.  
Then for $\delta\rightarrow 1+$,
\begin{eqnarray*}
\Bigr|\frac{1}{\sqrt{n}}\sum_{i=1}^n \left(U_i(\eta^*)-U_i(\hat{\eta})\right)\Bigr|
&\lesssim&\lambda_{\beta} \sqrt{n} c_{1\gamma}(n)+\lambda_{\gamma} \sqrt{n} c_{1\beta}(n)+\sqrt{n}c_2(n)^2.
\end{eqnarray*}
with probability tending to 1 under the sequence $P_n$.
In Section 1 of supplementary materials, we further demonstrate that (under regularity conditions stated in the same section), 
\begin{eqnarray*}
c_2(n)&=&\frac{\sqrt{(s_{\gamma}+s_{\beta})\log p}}{\sqrt{n}}\\
c_{1\gamma}(n)&=&s_{\gamma}\frac{\sqrt{\log p}}{\sqrt{n}}\\
c_{1\beta}(n)&=&s_{\beta}\frac{\sqrt{\log p}}{\sqrt{n}}.
\end{eqnarray*}
It follows that for $\delta\rightarrow 1+$,
\begin{eqnarray*}
\Bigr|\frac{1}{\sqrt{n}}\sum_{i=1}^n \left\{U_i(\eta^*)-U_i(\hat{\eta})\right\}\Bigr|
&=&O_{P_n}\left(\lambda_{\beta}s_{\gamma}\sqrt{\log p}\right)+O_{P_n}\left(\lambda_{\gamma}s_{\beta}\sqrt{\log p}\right)\\
&&+O_{P_n}\left(\frac{(s_{\gamma}+s_{\beta})\log p}{\sqrt{n}}\right).
\end{eqnarray*}
For default penalties satisfying $\lambda_{\gamma}=O\left(\sqrt{\frac{\log{p}}{n}}\right)$ and $\lambda_{\beta}=O\left(\sqrt{\frac{\log{p}}{n}}\right)$, we thus have that 
\begin{eqnarray*}
\Bigr|\frac{1}{\sqrt{n}}\sum_{i=1}^n \left\{U_i(\eta^*)-U_i(\hat{\eta})\right\}\Bigr|
&=&O_{P_n}\left(\frac{(s_{\gamma}+s_{\beta})\log p}{\sqrt{n}}\right),
\end{eqnarray*}
which converges to zero when $n\to \infty$, provided sufficient sparsity to ensure that $(s_{\gamma}+s_{\beta})\log p/\sqrt{n}\to 0$.
 $\Box$

The proof of the above proposition is instructive about the logic behind the above proposal. Repeating the 
same reasoning for the non-DR estimator $\tilde{\mu}$ with $U_i(\eta)=m(X_i;\beta)$ (and $\eta$ redefined as $\beta$), one finds that the term $\Bigr|\Bigr|\frac{1}{n}\sum_{i=1}^n \frac{\partial U_i}{\partial \beta} (\hat{\beta})\Bigr|\Bigr|_{\infty}$ is $O_{P_n}(1)$. It then follows that 
\begin{eqnarray*}
\biggr|\frac{1}{\sqrt{n}}\sum_{i=1}^n U_i(\eta^*)-U_i(\hat{\eta})\biggr|
&\lesssim& \sqrt{n} c_{1\beta}(n)+\sqrt{n}c_2(n)^2,
\end{eqnarray*}
with probability tending to 1 under the sequence $P_n$, in which the first term generally diverges to infinity. Likewise, repeating the 
above reasoning for the DR estimator $\hat{\mu}$ with nuisance parameter estimators obtained via standard lasso, one finds that the terms $\Bigr|\Bigr|\frac{1}{n}\sum_{i=1}^n \frac{\partial U_i}{\partial \beta} (\hat{\eta})\Bigr|\Bigr|_{\infty}$ and $\Bigr|\Bigr|\frac{1}{n}\sum_{i=1}^n \frac{\partial U_i}{\partial \gamma} (\hat{\eta})\Bigr|\Bigr|_{\infty}$ are $O_{P_n}(1)$, and not $o_{P_n}(1)$, unless both working models $\mathcal{A}$ and $\mathcal{B}$ are correctly specified in which case both gradients have expectation zero under the law $P_n$. Except under correct specification of both working models, the distribution of $\sqrt{n}(\hat{\mu}-\mu_0)$ is then generally complex and not well approximated by that of 
$n^{-1/2}\sum_{i=1}^n\left\{U_i(\eta^*)-\mu_0\right\}$. 


\subsection{Further properties}
\label{Section:2}

The procedure that we have proposed in Section \ref{proposal} was designed to make the empirical expectations
\begin{equation}\label{gradient}
\frac{1}{n}\sum_{i=1}^n  \frac{\partial }{\partial \gamma} U_i\left(\hat{\beta},\hat{\gamma}\right) \quad \mbox{\rm and} \quad \frac{1}{n}\sum_{i=1}^n \frac{\partial }{\partial \beta} U_i\left(\hat{\beta},\hat{\gamma}\right),\end{equation}
converge to zero. This has as a by-product that it makes the resulting estimator $\hat{\mu}$ insensitive to local changes in both nuisance parameters, provided that the sample size is sufficiently large. It is hence not entirely surprising that asymptotic inference based on $\hat{\mu}$ can ignore estimation of the nuisance parameters ${\beta}^*$ and ${\gamma}^*$, and that regularisation bias affecting these nuisance parameter estimators does not propagate into the estimator $\hat{\mu}$. \cite{farrell2015robust} also relies on this small bias property and finds it to hold regardless of the choice of nuisance parameter estimators, provided they both converge to their respective truths. This is because he implicitly relies on both 
models $\mathcal{A}$ and $\mathcal{B}$ being correctly specified, in which case the expectations (\ref{gradient}) converge to zero regardless of the choice of (consistent) estimator of the nuisance parameters. We have shown that this small bias property does not generally extend to contexts with model misspecification, unless when the nuisance parameters are estimated in accordance with the proposed procedure of Section \ref{proposal}.

In low-dimensional settings where the penalty parameters $\lambda_{\gamma}$ and $\lambda_{\beta}$ can be set to zero, the proposal reduces to the bias-reduced (BR) DR estimation procedure of \cite{vermeulen2015bias}. To gain insight into the behaviour of such procedures, we consider gross misspecification of the one-dimensional working models $\pi(X;\gamma)=\mbox{expit}(\gamma'(1,X_i))$ and $m(X;\beta)=\beta'(1,X_i)$ for two data-generating mechanisms (see the caption of Figures 1 and 2 for details); we deliberately focus on one-dimensional models so that the behaviour of the procedure can be clearly visualised. Figure 1 and 2 display the rescaled bias (i.e., sign(bias)$\sqrt{|\mathrm{bias}|}$) of the DR estimator for a range of nuisance parameter values. Upon contrasting both figures, one may see that the bias surface changes drastically as one of the data-generating models changes. The default DR estimator, which uses MLE for the nuisance parameters, therefore runs a great risk of ending up in a high bias zone. In contrast, the BR-DR estimator ends up in a saddle point of the bias surface.
The proposed BR-DR estimation principle thus locally minimises bias in certain directions of the nuisance parameters where the bias goes to plus infinity, and locally maximises it in other directions where the bias goes to minus infinity. Overall, much smaller biases of 2.34 and -9.4 are obtained for the BR-DR estimator in Figures 1 and 2, respectively, relative to the default DR estimator which has bias of 94.6 and -592; these calculations are based on a large sample of 100000 observations so as to approximate the asymptotic bias.
Moreover, even under misspecification of both working models, we would generally expect a more favourable bias of the BR-DR estimator than the Horvitz-Thompson (IPW) estimator 
\[\frac{1}{n}\sum_{i=1}^n \frac{A_iY_i}{\pi(X_i;\hat{\gamma})},\]
which is obtained upon setting $\beta$ to zero and $\gamma$ to the MLE. We would likewise generally expect more favourable bias than the imputation (IMP) estimator
\[\frac{1}{n}\sum_{i=1}^n A_iY_i+(1-A_i)m(X;\hat{\beta}),\]
which is obtained upon setting $\gamma$ to zero and $\beta$ to the solution to $0=\sum_{i=1}^n A_i(Y_i-\beta X_i)$. In Figures 1 and 2, we found the asymptotic bias to equal 71.5 and -633 for the IPW estimator, but to be merely 0.07 and 0.27 for the IMP estimator. This is partly due to happenstance: indeed, the BR-DR estimator would for instance have zero bias at a correctly specified propensity score model, unlike the imputation estimator. 

\begin{center}
{\bf Figures 1 and 2 about here.}
\end{center}

\section{Simulation study}
\label{Section:4}

In this section, we perform a simulation analysis to compare the performance of the proposed penalised bias-reduced estimator $\hat{\mu}_\mathrm{P-BR}$ with that of different estimators of a mean counterfactual outcome $\mu_0$. In particular, in subsection \ref{Sub:SimStudy1}, we detail the considered estimators of $\mu_0$. In subsection \ref{Sub:SimStudy2}, we describe the simulation scenarios for the models. In subsection \ref{Sub:SimStudy3}, we provide the discussion of the results. Finally, in subsection \ref{Sub:SimStudy4}, we numerically evaluate the behaviour of the proposed penalised bias-reduced estimator as the sample size increases, compared to competing approaches. 

\subsection{Considered Estimators and Settings}
\label{Sub:SimStudy1}

We denote nuisance parameters estimated through standard Maximum Likelihood Estimation and Ordinary Least Squares as $\hat{\eta}_\mathrm{MLE}=(\hat{\gamma}'_\mathrm{MLE},\hat{\beta}'_\mathrm{OLS})'$. We denote the nuisance parameters estimated through Lasso penalised Maximum Likelihood Estimation and Lasso penalised Least Squares as $\hat{\eta}_\mathrm{LASSO}=(\hat{\gamma}_\mathrm{LASSO}',\hat{\beta}_\mathrm{LASSO}')'$. Further, we denote nuisance parameters estimated through our proposed approach as $\hat{\eta}_\mathrm{P-BR}=(\hat{\gamma}_\mathrm{P-BR}',\hat{\beta}_\mathrm{P-BR}')'$. We additionally study the performance of the nuisance parameter estimators obtained through post-selection \citep{farrell2015robust} and double-selection techniques \citep{belloni2013honest, belloni2016post}. We denote these estimators as $\hat{\eta}_\mathrm{Post-LASSO}=(\hat{\gamma}'_\mathrm{Post-LASSO},\hat{\beta}_\mathrm{Post-LASSO}')'$ and $\hat{\eta}_\mathrm{DS-LASSO}=(\hat{\gamma}_\mathrm{DS-LASSO}',\hat{\beta}_\mathrm{DS-LASSO}')'$, respectively. In accordance with the double-selection procedure, we also evaluated a heuristic adaptation of the proposed procedure. In particular, applying the proposed bias-reduced DR estimation procedure resulted in the selection of covariate sets $X_{\hat{S}_\beta}$ in the outcome regression and $X_{\hat{S}_\gamma}$ in the propensity score regression. With $X$ set to $X_{\hat{S}}\equiv X_{\hat{S}_\beta}\cup X_{\hat{S}_\gamma}$, we next solved the following bias-reduced estimating equations with $\lambda$ set to zero:
\begin{eqnarray}
\label{DoubleSel1}
0&=&\sum_{i=1}^n \frac{\partial U_i(\eta)}{\partial\beta}=\sum_{i=1}^n \left\{1-\frac{A_i}{\pi(X_{i,\hat{S}},\gamma)}\right\}(1,X'_{i,\hat{S}})'\\
\label{DoubleSel2}
0&=&\sum_{i=1}^n \frac{\partial U_i(\eta)}{\partial\gamma}=-\sum_{i=1}^n \check{w}_iA_i\left\{Y_i-m(X_{i,\hat{S}},{\beta})\right\}(1,X'_{i,\hat{S}})', 
\end{eqnarray}
where 
\[\check{w}_i\equiv \frac{1-\pi(X_{i,\hat{S}},{\gamma})}{\pi(X_{i,\hat{S}},{\gamma})}.\]
The problem (\ref{DoubleSel1}) is computationally demanding under high-dimensional settings, however. Therefore, in order to solve it efficiently and guarantee numerical stability, we regularise the right hand side of (\ref{DoubleSel1}) through the penalty term $\lambda_{\gamma}\delta\hat{\gamma}^{\delta-1}$ with $\delta=2$.
This procedure may have the advantage that it makes the empirical analog of (\ref{gradient}) better satisfied in the sample and that it may reduce standard errors, but the disadvantage that the ridge penalisation induces another bias. We denote the resulting nuisance parameter estimator as $\hat{\eta}_\mathrm{DS-P-BR}=(\hat{\gamma}_\mathrm{DS-P-BR},\hat{\beta}_\mathrm{DS-P-BR})$.


We next consider the following estimators using the estimated nuisance parameters:
 \begin{enumerate}
 \item Regression Estimator: $\hat{\mu}_\mathrm{OR}(\hat{\beta})=\frac{1}{n} \displaystyle\sum_{i=1}^n m(X_i,\hat{\beta})$.
 \item Inverse-Propensity Weighting Estimators: $\hat{\mu}_\mathrm{IPTW}(\hat{\gamma})=\frac{1}{n} \displaystyle\sum_{i=1}^n {A_iY_i}{\pi^{-1}(X_i,\hat{\gamma})}$ and $\hat{\mu}_\mathrm{Pop-IPTW}(\hat{\gamma})=\displaystyle\sum_{i=1}^n {A_i Y_i \pi^{-1}(X_i,\hat{\gamma}_{MLE})}/\displaystyle\sum_{i=1}^n {A_i \pi^{-1}(X_i,\hat{\gamma}_{MLE})}$.
 \item DR estimators: $\hat{\mu}_{\mathrm{MLE}}=\hat{\mu}_{\mathrm{DR}}(\hat{\eta}_{MLE})$ (only when $n>p$), $\hat{\mu}_{\mathrm{LASSO}}=\hat{\mu}_{\mathrm{DR}}(\hat{\eta}_\mathrm{LASSO})$, $\hat{\mu}_{\mathrm{DS-LASSO}}=\hat{\mu}_{\mathrm{DR}}(\hat{\eta}_\mathrm{DS-LASSO})$, our proposed $\hat{\mu}_{\mathrm{P-BR}}=\hat{\mu}_{\mathrm{DR}}(\hat{\eta}_\mathrm{P-BR})$ and $\hat{\mu}_{\mathrm{DS-P-BR}}=\hat{\mu}_{\mathrm{DR}}(\hat{\eta}_\mathrm{DS-P-BR})$.  

\end{enumerate}  
 
In order to evaluate the performance of a given estimator $\hat{\mu}$, we consider the following measures: Monte Carlo Bias, Root Mean Square Error (RMSE), Median of Absolute Errors (MAE), Monte Carlo Standard Deviation (MCSD), Average of Sandwich Standard Errors (ASSE) and Monte Carlo Coverage (COV) of 95$\%$ confidence intervals. 

Note that several of the considered methods, including the proposed method, require the selection of the penalty parameter. Following the recommendation by \cite{belloni2016post} (see  \cite{meinshausen2006high} for a similar recommendation), we used the following choices:
\begin{eqnarray*}
\lambda_{\gamma} &=&\frac{1.1}{2\sqrt{n}}\Phi^{-1}\left(1-\frac{0.05}{\max(n, p\log n)}\right)\\
\lambda_{\beta}&=&\frac{1.1}{\sqrt{n}}\Phi^{-1}\left(1-\frac{0.05}{\max(n, p\log n)}\right),
\end{eqnarray*}
in our simulation study, in favour of low computational costs and in order to prevent biased standard errors as a result of ignoring the uncertainty in data-driven choices of $\lambda_{\beta}$ and $\lambda_{\beta}$.

\subsection{Simulation Scenarios}

In all simulation studies below, we generated $n$ mutually independent vectors $(X_i,A_i,Y_i)$, $i=1,...,n$. 
Here, ${X}_i=(X_{i,1}, ..., X_{i,p})$ is a mean zero multivariate normal covariate with covariance matrix $\Sigma$. We study the performance of the estimators for both, uncorrelated covariates (when $\Sigma=\textnormal{I}_{p\times p}$) and correlated covariates with covariance $\Sigma=[\sigma_{ij}]_{1\le i,j\le p}$ and $\sigma_{ij}=0.5^{|i-j|}$, for $i,j=1,...,p$. Note that in all cases the covariates have unit variance. Further, we let for each $i=1,...,n$, $A_i$ take on values 0 or 1 with $P(A_i=1|X_i)\equiv \pi_0({X}_i)$  and $Y_i$ be normally distributed with mean $m_0({X}_i)$ and unit variance, conditional on $X_i$  and $A_i=1$. In all studies, the simulated data were analysed using the following working models: $\pi(X,\beta)=\textnormal{expit}(\gamma_0+\displaystyle\sum_{i=1}^p\gamma_iX_i)$ and $m(X,\beta)=\beta_0+\displaystyle\sum_{i=1}^p\beta_iX_i$. For each data generating scenario, provided below, we conduct 1000 Monte Carlo runs with $n=200, \ p=40$ and $n=300, \ p=80$. 

In this section, we describe the results of two scenarios, and defer two additional simulation scenarios to the supplementary materials.
 
\label{Sub:SimStudy2}

\subsubsection{Scenario 1}
\label{Sub:Scenario1}
In the first scenario, we generated the data with $m_0({X})=\beta_0+cb'{X} $ and $\pi_0({X})=\textnormal{expit}(\gamma_0+g'{X})$, where $b\in \mathbb{R}^{p}$ and $g\in \mathbb{R}^{p}$ are defined as
\begin{eqnarray*}
b &=& (1, 1/2, 1/3, 1/4, 1/5, 0, 0, 0, 0, 0, 1, 1/2, 1/3, 1/4, 1/5, 0, 0,... , 0)\\
g &=& (1, 1/2, 1/3, 1/4, 1/5, 1/6, 1/7, 1/8, 1/9, 1/10, 0, 0, . . . , 0).
\end{eqnarray*}
We set $\beta_0=1, \gamma_0=0$ and $c=0.75$. These settings have been previously considered by \cite{belloni2013honest} and \cite{belloni2016post}. Finally, we also generated data with  $m({X})=X_{.,1}^2+b_{[2:p]}'X_{.,[2:p]}$ and $\pi({X})=\textnormal{expit}(X_{.,1}^2+g_{[2:p]}'X_{.,[2:p]})$ to evaluate the impact of model misspecification. Note that the target parameter $\mu_0=E(Y)$ is 1.


\subsubsection{Scenario 2}

In the second scenario, we use settings considered in \cite{kang2007demystifying} with $\pi_0({X})=\textnormal{expit}(-X_1+0.5X_2-0.25X_3-0.1X_4)$ and $m_0({X})=210+27.4X_1+13.7X_2+13.7X_3+13.7X_4$. The target parameter is $E(Y)=210$. The impact of model misspecification is evaluated via a linear outcome model and logistic propensity score model which are additive in the covariates $[M_1, M_2, M_3, X_4, ... X_p]$, where $M_1=\textnormal{exp}(X_1/2)$, $M_2=X_2/(1+\textnormal{exp}(X_1))+10$ and $M_3=(X_1X_3/25+0.6)^3$. 

\subsection{Discussion of Results}
\label{Sub:SimStudy3}
Tables 1 and 2 summarise the simulation results for $p=40$. We first consider the case where both models are correctly specified. As predicted by the theory (see the end of Section 2.3), the results for the data-adaptive estimators $\hat{\mu}_{\mathrm{OR}}(\hat{\beta}_\mathrm{LASSO})$ and  $\hat{\mu}_{\mathrm{Pop-IPTW}}(\hat{\gamma}_\mathrm{LASSO})$, which are not double-robust, 
show large bias and estimated standard errors that do not agree well with the empirical standard deviation. When both models are correctly specified, then using $\ell_1$-penalisation in combination with a DR estimator, as in $\hat{\mu}_{\mathrm{LASSO}}$, yields better performance because the first order terms in the Taylor expansion of Proposition \ref{MainProposition}  then have population mean converging to zero. The proposed estimator $\hat{\mu}_{\mathrm{P-BR}}$ sets these first order terms to zero, regardless of correct model specification, and this is observed to further reduce bias and improve mean squared error. 

In small sample sizes, the proposed estimators (just like other estimators based on penalisation) are subject to some residual bias. \cite{farrell2015robust} and \cite{belloni2016post} have proposed to eliminate some of this bias via the use of post-selection or double-selection, which is indeed seen to improve performance. This is generally also the case for the proposed procedure  
$\hat{\mu}_{\mathrm{DS-P-BR}}$, though not systematically because this procedure still uses $\ell_2$-penalisation for numerical stability in the fitting of the exposure model. As predicted by the theory, the proposed procedure $\hat{\mu}_{\mathrm{P-BR}}$ ensures that reasonable agreement between the estimated standard errors and the empirical standard deviation is obtained, even in settings with model misspecification. This is not guaranteed for the other DR estimators (with the exception of $\hat{\mu}_{\mathrm{DS-P-BR}}$), as is  most clearly seen in Scenario 2 (see Table 2), where misspecification of both models causes poor behaviour in the post-selection and double-selection procedures. 
 
 \begin{center}
{\bf Tables 1 and 2 about here.}
\end{center}

\subsection{Behaviour with increasing sample size}
\label{Sub:SimStudy4}

To evaluate the behaviour of the proposed estimator with increasing sample size, we reconsider the settings of Scenario 1 with $p=40$ and uncorrelated covariates, for sample sizes $n=\{200, 400, ..., 1000, 1500, 2000\}$. Table 3 provides the average measures over 1000 replications when both models are correctly specified and when the outcome model is misspecified. The results show that when both models are correctly specified, the Bias and RMSE of the proposed estimator $\hat{\mu}_\mathrm{P-BR}$ decrease and the coverage of the $95\%$ confidence interval improves with $n$. Moreover, $\hat{\mu}_\mathrm{P-BR}$ outperforms $\hat{\mu}_\mathrm{LASSO}$ throughout $n$ in terms of all measures. On the other hand, when the outcome model is misspecified, the Bias of $\hat{\mu}_\mathrm{P-BR}$ remains low over all considered sample sizes $n$. In contrast, we observe that when the outcome model is misspecified, the Bias of $\hat{\mu}_\mathrm{LASSO}$ surprisingly increases (in absolute value), resulting in a decreasing coverage with $n$. These results confirm the theory on the proposed estimator $\hat{\mu}_\mathrm{P-BR}$ when $n\to \infty$, and moreover suggest that also  the extended estimator $\hat{\mu}_\mathrm{DS-P-BR}$ has decreasing Bias and RMSE when $n$ increases.
 
 \begin{center}
{\bf Table 3 about here.}
\end{center}

\section{Illustration}
\label{Section:5}

In this section, we provide an empirical illustration of the proposed methodology on a real-data application. We study the effect of life expectancy (pseudo-exposure variable) on GDP growth (outcome variable). As in \cite{doppelhofer2009jointness}, we make use of World Bank data (\textnormal{\url{http://data.worldbank.org/}}) for 218 countries and dependencies and 9 covariates: population density (people per $\textnormal{km}^2$ of land area), total fertility rate (births per woman), exports of goods and services ($\%$ of GDP), imports of goods and services ($\%$ of GDP), Secure Internet servers (per 1 million people), land area ($\textnormal{km}^2$), mobile subscriptions (per 1000 people), mortality rate (per 1000 people under 5), unemployment ($\%$ of total labour force). After removing the observations with missing values, the final dataset consists of 152 observations. We consider data on life expectancy and covariates for the year 2013, and GDP growth for the year 2014. The constructed dataset includes 71 observations with low life expectancy below 73 years (i.e., roughly the median of life expectancy), coded $A=1$, and 81 observations with high life expectancy of at least 73 years, coded $A=0$. Our analysis here is intended only as an illustration, as it is a simplification of what is a more complex reality and therefore limited in the substantive conclusions that can be drawn. The causal effect of life expectancy on the GDP growth moreover forms a disputable topic in the literature \citep{acemoglu2007disease}.

In our analysis, we compare the methods considered in subsection \ref{Sub:SimStudy1} in both low and high-dimensional settings. In particular, for the first scenario, we consider only nine basic covariates. For the second scenario, in addition to the nine covariates, we also consider the squared and log transformations (in absolute values) of those covariates and all interactions between the basic ones. Thus, for the high-dimensional scenario, we consider 63 covariates. 

Table 4 summarises the estimated average treatment effects, sandwich estimators of the standard errors and 95$\%$ confidence intervals. It suggests that low life expectancy have negative effect on the GDP growth. It further shows that our proposed estimator $\hat{\mu}_{\mathrm{P-BR}}$ remains stable in terms of the standard errors when the dimension increases. In contrast, the performance of the estimator $\hat{\mu}_{\mathrm{MLE}}$ changes drastically as the number of covariates increases. 

We observe that, in the second scenario, the nuisance parameters estimated through our proposed approach contain several non-zero entries. In particular, 45 variables are selected using treated sub-sample and 42 variables are selected using untreated  sub-sample. Therefore, large number of selected covariates are considered for the double-selection equations (\ref{DoubleSel1}) and (\ref{DoubleSel2}). This produces estimation biases in the nuisance parameter estimator $\hat{\eta}_\mathrm{DS-P-BR}$. As a result, the standard error of the estimator $\hat{\mu}_{\mathrm{DS-P-BR}}$ increases significantly in the high-dimensional scenario.
   
 \begin{center}
{\bf Table 4 about here.}
\end{center}

\section{Discussion}
\label{Section:6}

Plug-in estimators based on data-adaptive high-dimensional model fits are well known to exhibit poor behaviour with non-standard asymptotic distribution \citep{pfanzagll982, van2011targeted}. Double-robust plug-in estimators have been shown to be much less sensitive to this when all working models on which they are based are correctly specified (or estimators for them converge to the truth) \citep{farrell2015robust}. In this paper, we have shown that this continues to be true under model misspecification when so-called penalised bias-reduced double-robust estimators are used. These estimators can be viewed as a penalised extension of recently introduced bias-reduced DR estimators, which use special nuisance parameter estimators that are designed to minimise - or at least stabilise - the squared first-order bias of the DR estimator, while shrinking the non-significant coefficients of the nuisance parameters towards zero. Our results thus generalise those in \cite{belloni2013honest}, \cite{farrell2015robust} and \cite{belloni2016post} to allow for model misspecification. Through extensive simulation studies, we have demonstrated that the proposed approach performs favourably compared to other DR estimators even when one of the models are misspecified. The empirical data analysis further confirmed the stability of our estimator of the average treatment effect in terms of the standard errors as the dimension of the covariates increases. We did not yet consider settings with $p>n$ in view of the computational difficulty of minimising the objective function in that case, and plan to address this in future work. 

We have focussed our numerical results on lasso or $\ell_1$-norm penalisation, even though it readily generalises to other (possibly non-convex) penalisation techniques. It remains to be seen how it performs in combination with other choices of penalty. Our theory, like that in \cite{farrell2015robust} and \cite{belloni2016post}, was also developed for prespecified penalty parameters, although the calibration of penalty parameters is likely to improve results. In further work, we will evaluate whether our theory can be adapted to incorporate data-adaptive choices of penalty parameters, e.g. based on cross-validation. We conjecture (and have confirmed in limited numerical studies - not reported) that our proposal may, by construction, deliver DR estimators which have limited sensitivity to the chosen regularisation procedure (e.g. to the choice of penalty used for estimating the nuisance parameters), as well as to mild misspecification of both models $\mathcal{A}$ and $\mathcal{B}$. 

We have explored the use of ad-hoc debiasing steps based on post-lasso, and found mixed success with the proposed approach. This is likely related to the fact that the considered double-selection procedure sometimes leads to the selection of many covariates, and moreover to the use of a ridge penalty in order to guarantee numerical stability of the optimisation procedure. In future work, we will consider the potential to de-bias the solutions to the proposed estimating equations (\ref{EstEq1})-(\ref{EstEq2}) along the lines of \cite{zhang2014confidence}, \cite{van2014asymptotically}. 
 
\cite{belloni2016post} show that the use of sample splitting may lead to less stringent sparsity conditions. In particular, they find that $\sqrt{s_{\gamma}s_{\beta}} \log(p)/\sqrt{n}$ converging to zero is sufficient to guarantee uniformly valid confidence intervals when both models are correctly specified. This is attractive as it enables one model to be dense, so long as the other is known to be sparse, as is typically the case in the context of randomised experiments. In contrast, we require that 
$\lambda_{\beta} \sqrt{n} c_{1\gamma}(n)+\lambda_{\gamma} \sqrt{n} c_{1\beta}(n)+\sqrt{n}c_2(n)^2$ converges to zero. In simple randomised experiments, $s_{\gamma}=0$ so that fast convergence rates of $\hat{\gamma}$ (i.e., $c_{1\gamma}(n)$ converging to zero at a fast rate) are attainable even when $\lambda_{\gamma}$ is very small. This creates potential for making $\lambda_{\beta} \sqrt{n} c_{1\gamma}(n)+\lambda_{\gamma} \sqrt{n} c_{1\beta}(n)$ converge to zero in the context of randomised experiments, even when dense outcome models are used. To what extent and under what conditions this is achievable, will be investigated in future work. We furthermore plan to evaluate whether stronger results are achievable with sample splitting.

Finally, at a more general level, our results indicate that the choice of nuisance parameter estimators can matter a lot in settings with model misspecification, and that important benefits may be achievable via the choice of special nuisance parameter estimators. We hope that this work will not only help to achieve inferences with greater validity in the presence of variable selection, but moreover stimulate research on more general statistical learning procedures for the working models indexing a DR estimator, targeted towards achieving reliable inferences even when the usual modelling or sparsity assumptions are not met.

\bibliography{mybib}

\newpage

\begin{figure}[h!]
\centering
\includegraphics[width=\textwidth]{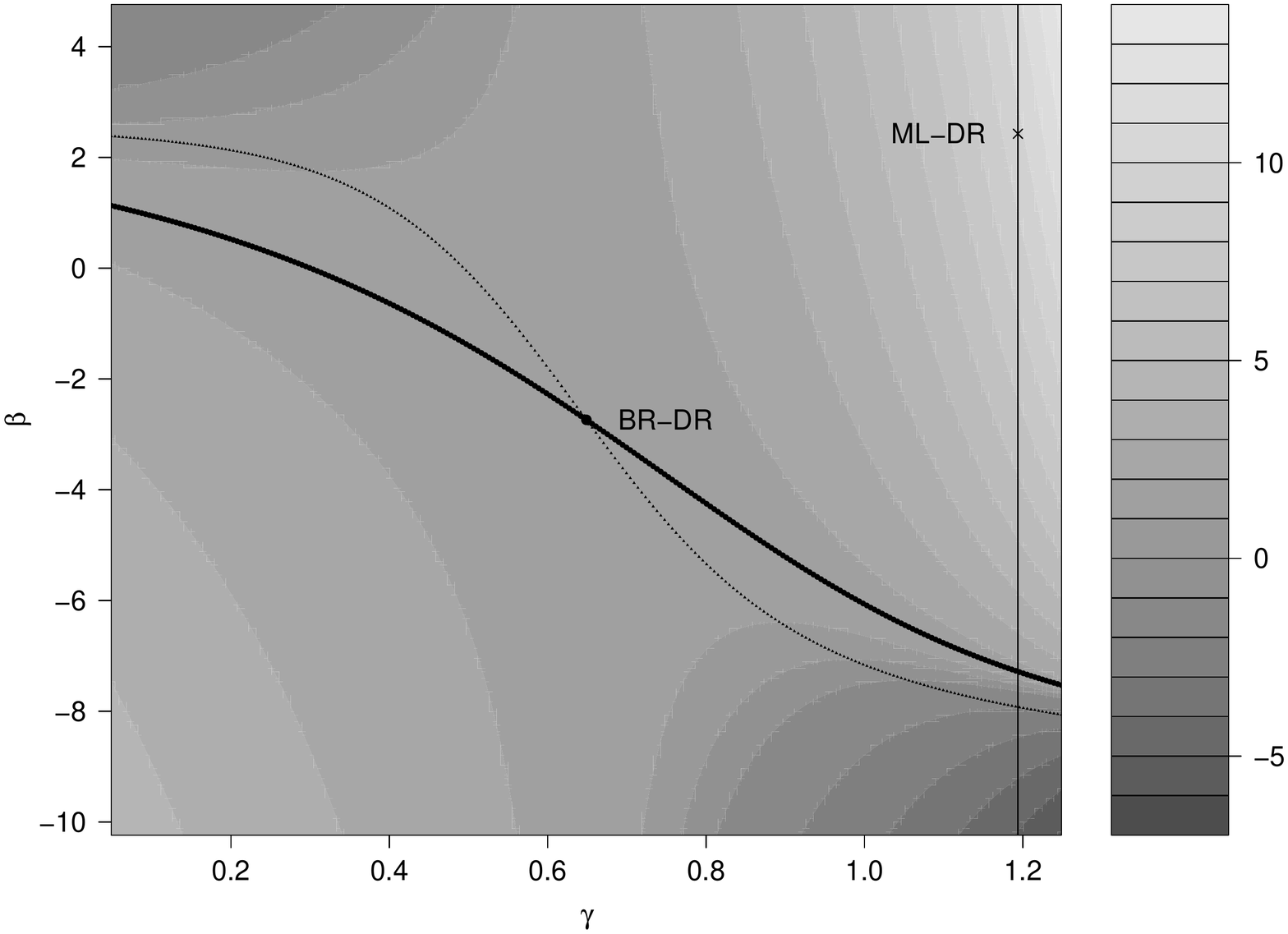}
\caption{Rescaled bias (sign(bias)$\sqrt{|\mathrm{bias}|}$) of the DR estimator of $E\{Y(1)\}$ in function of the nuisance parameter values $\gamma$ and $\beta$ under the following data-generating model: $X=(3-V)/SD(3-V)$ with $V$ a Gamma variate with scale and shape 1, $P(A=1|X)=\mathrm{expit}(-1+X^2)$ and $Y\sim N(X^2,1)$. BR: bias-reduced estimator; MLE: maximum likelihood estimator; MLE-BR: bias-reduced estimator of $\beta$, conditional on maximum likelihood estimator of $\gamma$. Dotted line shows the bias-reduced estimator of $\beta$, conditional on $\gamma$.}
\end{figure}
\newpage

\begin{figure}[h!]
\centering
\includegraphics[width=\textwidth]{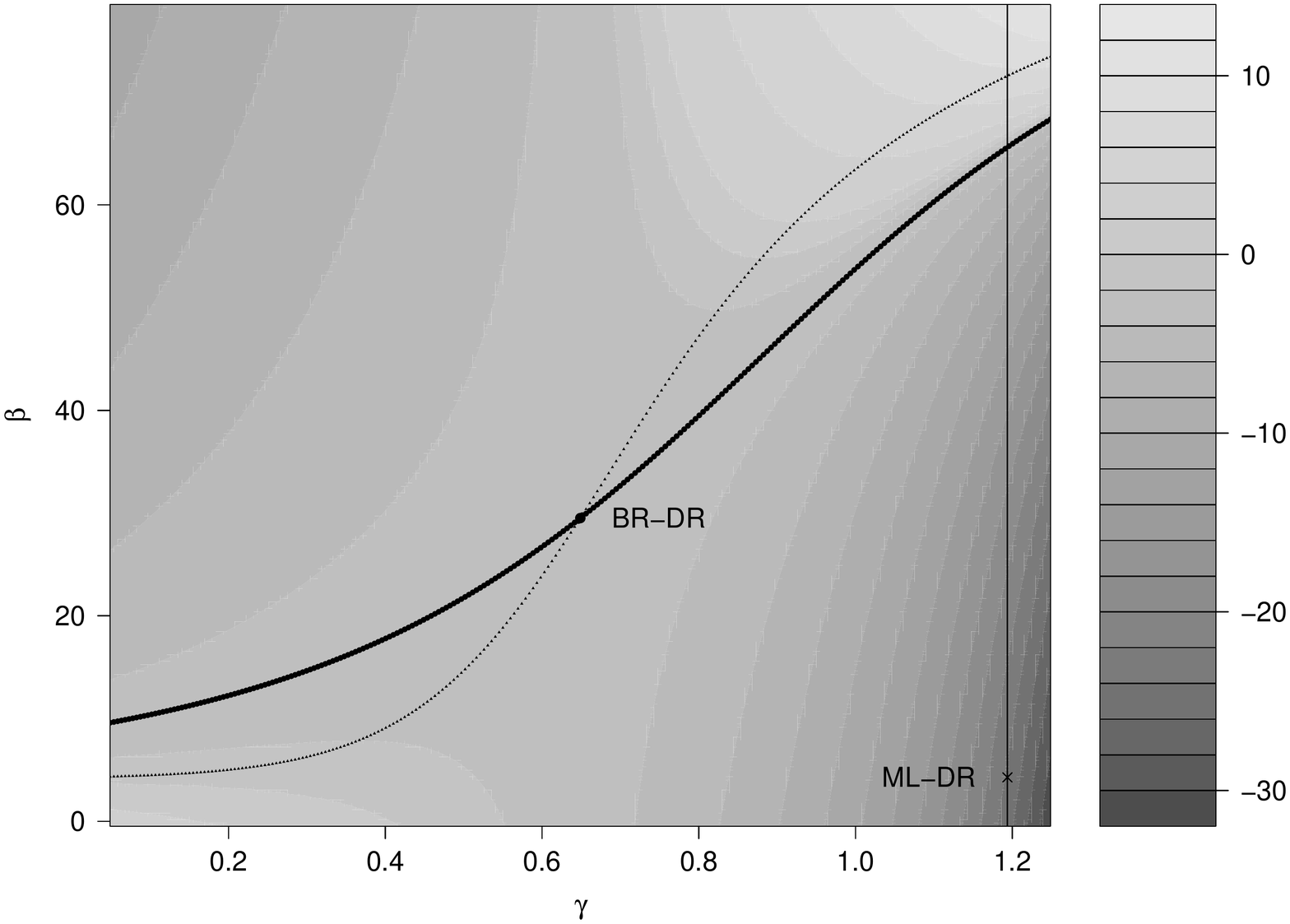}
\caption{Rescaled bias (sign(bias)$\sqrt{|\mathrm{bias}|}$) of the DR estimator of $E\{Y(1)\}$ in function of the nuisance parameter values $\gamma$ and $\beta$ under the following data-generating model: $X=(3-V)/SD(3-V)$ with $V$ a Gamma variate with scale and shape 1, $P(A=1|X)=\mathrm{expit}(-1+X^2)$ and $Y\sim N(X^3-X^2,1)$. BR: bias-reduced estimator; MLE: maximum likelihood estimator; MLE-BR: bias-reduced estimator of $\beta$, conditional on maximum likelihood estimator of $\gamma$. Dotted line shows the bias-reduced estimator of $\beta$, conditional on $\gamma$.}
\end{figure}

\newpage

{\renewcommand{\arraystretch}{1.1}
 \begin{table}[h!]
   \centering
  \captionsetup{justification=centering}
   \caption{Simulation results based on 1000 replications, Scenario 1, $p=40, n=200$.}
   \label{Scenario2Table1}
    \resizebox{14cm}{!}{   \begin{tabular}{l  r c r  r  r r| r c  r   r  r r }
   \hline
\hline

   Estimator & Bias  & RMSE & MAE  & MCSD  & ASSE & COV & Bias  & RMSE & MAE  & MCSD  & ASSE & COV \\  \hline
             &      &      Uncorrelated  &     &       &      &                  &      &  Correlated      &     &       &      &       \\ \hline
 \textbf{ OR correct }         &      &        &     &       &      &                  &      &       &     &       &      &       \\ 
 \textbf{ PS correct }         &      &        &     &       &      &                  &      &       &     &       &      &       \\ 
$\hat{\mu}_\mathrm{OR}(\hat{\beta}_\mathrm{OLS})$	&	0.001	&	0.158	&	0.110	&	0.158	&	0.104	&	0.797	&	0.0003	&	0.185	&	0.121	&	0.185	&	0.132	&	0.832	\\ 
$\hat{\mu}_\mathrm{Pop-IPTW}(\hat{\gamma}_\mathrm{MLE})$	&	0.006	&	0.342	&	0.160	&	0.342	&	0.255 &  0.908	& 0.053	&	0.541	&	0.287	&	0.539	&	0.330   &   0.821	\\ 
$\hat{\mu}_\mathrm{OR}(\hat{\beta}_\mathrm{LASSO})$  	&	0.249	&	0.291	&	0.246	&	0.151	&	0.047	&	0.141	&	0.302	&	0.348	&	0.308	&	0.173	&	0.080	&	0.214	\\ 
$\hat{\mu}_\mathrm{Pop-IPTW}(\hat{\gamma}_\mathrm{LASSO})$ 	&	0.354	&	0.386	&	0.353	&	0.153	&	0.158 &  0.397		&	0.562	&	0.590	&	0.567	&	0.181&0.190 &     0.163\\ 
$\hat{\mu}_\mathrm{MLE}$	&	-0.006	&	0.318	&	0.122	&	0.318	&	0.182	&	0.916 &	-0.026	&	0.480	&	0.146	&	0.479	&	0.232	&	0.905	\\
 $\hat{\mu}_\mathrm{LASSO}$	&	0.222	&	0.268	&	0.222	&	0.150	&	0.136	&	0.610 &	0.252	&	0.306	&	0.259	&	0.173	&	0.149	&	0.577	\\ 
$\hat{\mu}_\mathrm{DS-LASSO}$	&	0.080	&	0.181	&	0.124	&	0.162	&	0.148	&	0.872	&	0.025	&	0.199	&	0.131	&	0.197	&	0.184	&	0.934	\\
$\hat{\mu}_\mathrm{Post-LASSO}$ 	&	0.081	&	0.180	&	0.123	&	0.160	&	0.143	&	0.864	&	0.028	&	0.187	&	0.129	&	0.185	&	0.177	&	0.933	\\ 
$\hat{\mu}_\mathrm{P-BR}$	&	0.144	&	0.211	&	0.151	&	0.153	&	0.135	&	0.765	&	0.148	&	0.239	&	0.167	&	0.188	&	0.151	&	0.757	\\ 
 $\hat{\mu}_\mathrm{DS-P-BR}$	&	0.032	&	0.162	&	0.113	&	0.158	&	0.130	&	0.875	&	0.019	&	0.199	&	0.134	&	0.198	&	0.150	&	0.870	\\  \hline 
 
 \textbf{ OR incorrect }         &      &        &     &       &      &                  &      &       &     &       &      &       \\ 
 \textbf{ PS correct }         &      &        &     &       &      &                  &      &       &     &       &      &       \\ 

$\hat{\mu}_\mathrm{OR}(\hat{\beta}_\mathrm{OLS})$	&	-0.308	&	0.391	&	0.315	&	0.240	&	0.124	&	0.366 &	-0.451	&	0.524	&	0.454	&	0.267	&	0.154	&	0.283	\\
$\hat{\mu}_\mathrm{Pop-IPTW}(\hat{\gamma}_\mathrm{MLE})$	&	-0.033	&	0.424	&	0.197	&	0.423	&	0.295 &  0.921	&	-0.007	&	0.578	&	0.236	&	0.578	&	0.340 & 0.920	\\
$\hat{\mu}_\mathrm{OR}(\hat{\beta}_\mathrm{LASSO})$  	&	-0.067	&	0.215	&	0.149	&	0.204	&	0.055	&	0.365&	-0.152	&	0.273	&	0.191	&	0.226	&	0.093	&	0.473	\\
$\hat{\mu}_\mathrm{Pop-IPTW}(\hat{\gamma}_\mathrm{LASSO})$ 	&	0.082	&	0.214	&	0.147	&	0.198	&	0.192 &  0.937&	0.230	&	0.316	&	0.240	&	0.217	&	0.215 &     0.819	\\ 
$\hat{\mu}_\mathrm{MLE}$	&	-0.129	&	0.489	&	0.268	&	0.471	&	0.305	&	0.780	&	-0.178	&	2.149	&	0.377	&	2.142	&	0.502	&	0.670	\\ 
 $\hat{\mu}_\mathrm{LASSO}$	&	-0.074	&	0.219	&	0.149	&	0.205	&	0.174	&	0.877	&	-0.170	&	0.284	&	0.204	&	0.227	&	0.183	&	0.777	\\ 
$\hat{\mu}_\mathrm{DS-LASSO}$	&	-0.007	&	0.323	&	0.181	&	0.323	&	0.261	&	0.890	&	-0.103	&	0.495	&	0.271	&	0.484	&	0.343	&	0.813\\
$\hat{\mu}_\mathrm{Post-LASSO}$ 	&	0.001	&	0.306	&	0.185	&	0.306	&	0.256	&	0.909	&	-0.085	&	0.500	&	0.255	&	0.493	&	0.345	&	0.831	\\ 

 $\hat{\mu}_\mathrm{P-BR}$	&	-0.010	&	0.201	&	0.141	&	0.201	&	0.167	&	0.898	&	-0.046	&	0.233	&	0.162	&	0.228	&	0.173	&	0.842	\\ 
 $\hat{\mu}_\mathrm{DS-P-BR}$	&	-0.132	&	0.262	&	0.182	&	0.226	&	0.160	&	0.749&	-0.194	&	0.331	&	0.222	&	0.268	&	0.171	&	0.693	\\ \hline

 \textbf{ OR correct }         &      &        &     &       &      &                  &      &       &     &       &      &       \\ 
 \textbf{ PS incorrect }         &      &        &     &       &      &                  &      &       &     &       &      &       \\ 
$\hat{\mu}_\mathrm{OR}(\hat{\beta}_\mathrm{OLS})$	&	-0.0008	&	0.133	&	0.092	&	0.133	&	0.099	&	0.857	&	-0.002	&	0.156	&	0.108	&	0.156	&	0.129	&	0.899	\\
$\hat{\mu}_\mathrm{Pop-IPTW}(\hat{\gamma}_\mathrm{MLE})$	&	-0.005	&	0.183	&	0.103	&	0.183	&	0.173 &  0.977	&	-0.022	&	0.258	&	0.128	&	0.257	&	0.238 & 0.971	\\ 
$\hat{\mu}_\mathrm{OR}(\hat{\beta}_\mathrm{LASSO})$  	&	0.077	&	0.152	&	0.106	&	0.130	&	0.052	&	0.469	&	0.095	&	0.180	&	0.131	&	0.153	&	0.087	&	0.641	\\
$\hat{\mu}_\mathrm{Pop-IPTW}(\hat{\gamma}_\mathrm{LASSO})$ 	&	0.093	&	0.169	&	0.119	&	0.141	&	0.145 & 0.914	&	0.229	&	0.286	&	0.239	&	0.171	&	0.179 & 0.777	\\ 
$\hat{\mu}_\mathrm{MLE}$	&	0.004	&	0.230	&	0.096	&	0.231	&	0.138	&	0.937	&	$10^{-5}$	&	0.171	&	0.114	&	0.171	&	0.160	&	0.938	\\ 
 $\hat{\mu}_\mathrm{LASSO}$	&	0.077	&	0.151	&	0.106	&	0.130	&	0.131	&	0.912	&	0.090	&	0.177	&	0.126	&	0.153	&	0.152	&	0.908	\\ 
$\hat{\mu}_\mathrm{DS-LASSO}$	&	0.036	&	0.136	&	0.095	&	0.131	&	0.127	&	0.938	&	0.006	&	0.152	&	0.103	&	0.152	&	0.153	&	0.954	\\ 
$\hat{\mu}_\mathrm{Post-LASSO}$ 	&	0.036	&	0.136	&	0.095	&	0.131	&	0.126	&	0.935	&	0.005	&	0.151	&	0.103	&	0.151	&	0.151	&	0.953	\\ 

$\hat{\mu}_\mathrm{P-BR}$	&	0.068	&	0.147	&	0.104	&	0.130	&	0.144	&	0.950	&	0.062	&	0.165	&	0.114	&	0.152	&	0.165	&	0.954	\\ 
 $\hat{\mu}_\mathrm{DS-P-BR}$	&	0.018	&	0.131	&	0.093	&	0.130	&	0.132	&	0.959	&	-0.0009	&	0.153	&	0.107	&	0.153	&	0.154	&	0.948	\\ 
\hline
 \textbf{ OR incorrect }         &      &        &     &       &      &                  &      &       &     &       &      &       \\ 
 \textbf{ PS incorrect }         &      &        &     &       &      &                  &      &       &     &       &      &       \\ 
$\hat{\mu}_\mathrm{OR}(\hat{\beta}_\mathrm{OLS})$	&	0.321	&	0.382	&	0.311	&	0.208	&	0.104	&	0.302	&	0.347	&	0.409	&	0.338	&	0.218	&	0.123	&	0.310	\\ 
$\hat{\mu}_\mathrm{Pop-IPTW}(\hat{\gamma}_\mathrm{MLE})$	&	0.329	&	0.418	&	0.313	&	0.258	&	0.218 &  0.674	&	0.380	&	0.485	&	0.371	&	0.301	&	0.250 & 0.640 	\\ 
$\hat{\mu}_\mathrm{OR}(\hat{\beta}_\mathrm{LASSO})$  	&	0.376	&	0.421	&	0.367	&	0.188	&	0.041	&	0.053	&	0.421	&	0.466	&	0.416	&	0.198	&	0.067	&	0.077	\\ 
$\hat{\mu}_\mathrm{Pop-IPTW}(\hat{\gamma}_\mathrm{LASSO})$ 	&	0.389	&	0.433	&	0.380	&	0.190	&	0.184 &  0.446	&	0.490	&	0.530	&	0.487	&	0.202	&	0.201 & 0.317	\\ 
$\hat{\mu}_\mathrm{MLE}$	&	0.359	&	1.124	&	0.319	&	1.066	&	0.230	&	0.598	&	0.382	&	0.581	&	0.362	&	0.437	&	0.230	&	0.575	\\ 
 $\hat{\mu}_\mathrm{LASSO}$	&	0.376	&	0.420	&	0.367	&	0.188	&	0.177	&	0.446	&	0.417	&	0.462	&	0.413	&	0.199	&	0.188	&	0.397	\\ 
$\hat{\mu}_\mathrm{DS-LASSO}$	&	0.352	&	0.404	&	0.338	&	0.198	&	0.176	&	0.501	&	0.383	&	0.442	&	0.371	&	0.221	&	0.204	&	0.529	\\
$\hat{\mu}_\mathrm{Post-LASSO}$ 	&	0.348	&	0.401	&	0.335	&	0.197	&	0.173	&	0.496	&	0.370	&	0.430	&	0.361	&	0.219	&	0.197	&	0.529	\\ 
 $\hat{\mu}_\mathrm{P-BR}$	&	0.370	&	0.416	&	0.363	&	0.189	&	0.199	&	0.558	&	0.411	&	0.457	&	0.406	&	0.201	&	0.211	&	0.509	\\ 
 $\hat{\mu}_\mathrm{DS-P-BR}$	&	0.338	&	0.394	&	0.326	&	0.202	&	0.187	&	0.583	&	0.373	&	0.431	&	0.375	&	0.215	&	0.199	&	0.534	\\ 
 \hline \hline

 \end{tabular}
   }
  
   \end{table}
   }
 
  \scriptsize{ \textnormal{NOTE: Bias: Monte Carlo Bias, RMSE: Root Mean Square Error, MAE: Median of Absolute Errors, MCSD: Monte Carlo Standard Deviation, COV: coverage of $95\%$ confidence intervals, OR: Outcome Regression, PS: Propensity Score. For the settings OR correct, PS correct, correlated covariates and OR incorrect, PS correct, correlated covariates, no convergence was attained for $\hat{\mu}_\mathrm{P-BR}$ in one run, for $\hat{\mu}_\mathrm{DS-P-BR}$ in four runs out of 1000. } }
  
  \normalsize

  \newpage
  
  {\renewcommand{\arraystretch}{1.3}
      \begin{table}[h!]
   \centering
  \captionsetup{justification=centering}
   \caption{Simulation results based on 1000 replications, Scenario 2, $p=40, n=200$.}
   \label{Scenario4Table1}
    \resizebox{15cm}{!}{   \begin{tabular}{l  r c r  r  r r| r c  r   r  r r }
   \hline
\hline

   Estimator & Bias  & RMSE & MAE  & MCSD  & ASSE & COV & Bias  & RMSE & MAE  & MCSD  & ASSE & COV \\  \hline
              &      &      Uncorrelated  &     &       &      &                  &      &  Correlated      &     &       &      &       \\ \hline
  \textbf{ OR correct }         &      &        &     &       &      &                  &      &       &     &       &      &       \\ 
  \textbf{ PS correct }         &      &        &     &       &      &                  &      &       &     &       &      &       \\ 
$\hat{\mu}_\mathrm{OR}(\hat{\beta}_\mathrm{OLS})$	        &   0.089 & 2.520 & 1.668 & 2.520 &  2.566 & 0.952 &	0.122  &   3.478 & 2.404 & 3.478 & 3.498 & 0.954 \\
$\hat{\mu}_\mathrm{Pop-IPTW}(\hat{\gamma}_\mathrm{MLE})$	&	0.082 & 6.900 & 3.470 & 6.903 & 5.545 & 0.939  &   -0.235 &   7.282 & 4.140 & 7.282 & 7.181 & 0.959 	\\
$\hat{\mu}_\mathrm{OR}(\hat{\beta}_\mathrm{LASSO})$  	    &  -0.022 & 2.512 & 1.679 & 2.513 & 2.528 & 0.947	&	0.004  &   3.471 & 2.350 & 3.472 & 3.468 & 0.951\\
$\hat{\mu}_\mathrm{Pop-IPTW}(\hat{\gamma}_\mathrm{LASSO})$ 	&	-7.259 & 7.852 & 7.214 & 2.994 & 3.552 & 0.461  &	-10.76 &   11.50 & 10.69 & 4.079 & 4.805 & 0.374 \\
$\hat{\mu}_\mathrm{MLE}$	&	0.100  & 2.531 & 1.691 & 2.530 & 2.573 & 0.950	&	0.117  &   3.483 & 2.387 & 3.482 & 3.500 & 0.953\\
 $\hat{\mu}_\mathrm{LASSO}$	&	0.005  & 2.513 & 1.680 & 2.514 & 2.563 & 0.955 	&	0.023  &   3.471 & 2.368 & 3.473 & 3.495 & 0.955 \\
$\hat{\mu}_\mathrm{DS-LASSO}$	&	0.087 & 2.518 & 1.667 & 2.518 & 2.568 & 0.952	&	0.112  &   3.475 & 2.379 & 3.475 & 3.498 & 0.952 \\
$\hat{\mu}_\mathrm{Post-LASSO}$ 	&	0.085 & 2.517 & 1.682 & 2.517 & 2.568 & 0.952 &	0.111  &   3.474 & 2.377 & 3.474 & 3.498 &  0.953\\ 
 $\hat{\mu}_\mathrm{P-BR}$	&	0.038 &  2.517 & 1.690 & 2.518 & 2.562 & 0.956&	0.069  &   3.475 & 2.372 & 3.476 & 3.495 & 0.951 	\\
 $\hat{\mu}_\mathrm{DS-P-BR}$	&	0.082 & 2.514 & 1.698 & 2.514 & 2.566 & 0.957  &	0.111  &   3.475 & 2.402 & 3.475 & 3.497 & 0.953	\\
 \hline

  \textbf{ OR incorrect }         &      &        &     &       &      &                  &      &       &     &       &      &       \\ 
  \textbf{ PS incorrect }         &      &        &     &       &      &                  &      &       &     &       &      &       \\ 
$\hat{\mu}_\mathrm{OR}(\hat{\beta}_\mathrm{OLS})$	&	0.723	&	3.645	&	2.539	&	3.574	&	2.801	&	0.878	&	0.344	&	4.016	&	2.799	&	4.003	&	3.591	&	0.929	\\
$\hat{\mu}_\mathrm{Pop-IPTW}(\hat{\gamma}_\mathrm{MLE})$	&	2.104	&	12.65	&	4.026	&	12.48	&	6.940	&	0.925 &	3.095	&	14.21	&	4.882	&	13.88	&	8.827	&	0.953	\\
$\hat{\mu}_\mathrm{OR}(\hat{\beta}_\mathrm{LASSO})$  	&	0.580	&	3.513	&	2.474	&	3.466	&	2.714	&	0.882	&	0.187	&	3.933	&	2.737	&	3.931	&	3.529	&	0.925\\
$\hat{\mu}_\mathrm{Pop-IPTW}(\hat{\gamma}_\mathrm{LASSO})$ 	&	-8.249	&	8.810	&	8.251	&	3.095	&	3.584	&	0.351 &	-11.84	&	12.58	&	11.77	&	4.241	&	4.837	&	0.280	\\
$\hat{\mu}_\mathrm{MLE}$	&	-6.832	&	68.59	&	3.012	&	68.28	&	9.740	&	0.936	&	-2.279	&	16.84	&	3.301	&	16.70	&	5.498	&	0.940\\
 $\hat{\mu}_\mathrm{LASSO}$	&	0.550	&	3.513	&	2.470	&	3.472	&	2.980	&	0.916	&	0.185	&	3.934	&	2.740	&	3.931	&	3.699	&	0.939	\\
$\hat{\mu}_\mathrm{DS-LASSO}$	&	-5.369	&	48.38	&	2.991	&	48.11	&	8.148	&	0.940	&	-2.521	&	18.78	&	3.132	&	18.62	&	5.551	&	0.936	\\
$\hat{\mu}_\mathrm{Post-LASSO}$ 	&	-2.709	&	18.04	&	2.853	&	17.84	&	5.362	&	0.925	&	-0.741	&	5.555	&	2.849	&	5.508	&	4.228	&	0.946	\\
$\hat{\mu}_\mathrm{P-BR}$	&	-0.086	&	3.398	&	2.391	&	3.399	&	2.952	&	0.909	&	-0.085	&	3.884	&	2.654	&	3.885	&	3.695	&	0.936	\\
 $\hat{\mu}_\mathrm{DS-P-BR}$	&	0.117	&	3.491	&	2.507	&	3.491	&	2.974	&	0.907	&	0.034	&	3.980	&	2.768	&	3.982	&	3.707	&	0.932	\\

 \hline\hline

   \end{tabular}
   }
   \end{table}
   }
  
  \scriptsize{ \textnormal{NOTE: Bias: Monte Carlo Bias, RMSE: Root Mean Square Error, MAE: Median of Absolute Errors, MCSD: Monte Carlo Standard Deviation, COV: coverage of $95\%$ confidence intervals, OR: Outcome Regression, PS: Propensity Score. } }
  
  \normalsize
  
\newpage
{\renewcommand{\arraystretch}{1.2}
    \begin{table}[h!]
  \centering
   \captionsetup{justification=centering}
  \caption{Bias, Root Mean Squared Error (RMSE) and coverage (COV) of $95\%$ confidence intervals based on 1000 replications in Scenario 1 for $p=40$ and different values of $n$.}
  \resizebox{14cm}{!}{
   \begin{tabular}{l  l r  r r r r r r r r}
   \hline
   OR correct & PS correct & & & & & & \\
   \hline
   Estimator & Measure  & $n=200$ & $n=400$ & $n=600$ & $n=800$ & $n=1000$ & $n=1500$ & $n=2000$  \\ \hline
   $\hat{\mu}_{\mathrm{P-BR}}$ & Bias & 0.144 & 0.098 & 0.079 & 0.063 & 0.052 & 0.039 & 0.029\\
   							& RMSE & 0.211 & 0.145 & 0.118 & 0.099 & 0.088 & 0.066 & 0.056\\
   							& COV & 0.765 & 0.794 & 0.815 & 0.826 & 0.835 & 0.870 &  0.869\\ \hline
   $\hat{\mu}_{\mathrm{LASSO}}$ & Bias & 0.222 & 0.168 & 0.142 & 0.122 & 0.107 & 0.086 & 0.070\\ 
     						& RMSE & 0.268 & 0.197 & 0.166 & 0.142 & 0.127 & 0.100 & 0.084\\
   							& COV & 0.610 & 0.575 & 0.529 & 0.541 & 0.549 & 0.574 & 0.608\\ \hline 
   							
     $\hat{\mu}_{\mathrm{MLE}}$ & Bias & -0.006 & -0.002 & 0.001 & 0.0007 & 0.001 & 0.002 &0.0001 \\ 
     						& RMSE & 0.318 & 0.125 & 0.096 & 0.079 & 0.075 & 0.056 & 0.050\\ 
   							& COV & 0.916 & 0.937 & 0.940 & 0.946 & 0.940 & 0.954 & 0.947\\ \hline
   							
   	 $\hat{\mu}_{\mathrm{DS-P-BR}}$ & Bias & 0.032 & 0.012  & 0.010 & 0.004 & 0.004 & 0.004 & 0.001\\ 
     						& RMSE       & 0.162 & 0.111  & 0.090 & 0.076  & 0.071 & 0.053 & 0.048\\
   							& COV   & 0.875 & 0.906  & 0.919 & 0.917  & 0.911 & 0.943 & 0.927\\ \hline
   							
   	   	 $\hat{\mu}_{\mathrm{DS-LASSO}}$ & Bias & 0.080 & 0.041  & 0.026 & 0.015 & 0.011 & 0.005 & 0.001\\ 
     						& RMSE       & 0.181 & 0.119  & 0.094 & 0.077  & 0.074 & 0.055 & 0.048\\ 
   							& COV   & 0.872 & 0.921  & 0.930 & 0.935  & 0.931 & 0.953 & 0.953\\ \hline\\
   	 OR incorrect & PS correct & & & & & & \\
   \hline						
   	 Estimator & Measure  & $n=200$ & $n=400$ & $n=600$ & $n=800$ & $n=1000$ & $n=1500$ & $n=2000$  \\ \hline
   $\hat{\mu}_{\mathrm{P-BR}}$ & Bias & -0.010 & -0.007 & -0.005 & -0.006 &  -0.007 & -0.003 & -0.008 \\ 
   							& RMSE & 0.201 & 0.146 & 0.125 & 0.111 & 0.100 & 0.080 & 0.073\\ 
   							& COV & 0.898 & 0.899 & 0.903 & 0.894 & 0.889 & 0.909 & 0.894 \\ \hline
   							
   $\hat{\mu}_{\mathrm{LASSO}}$ & Bias & -0.074 & -0.093 & -0.101 & -0.111 & -0.117 & -0.115 & -0.123\\ 
     						& RMSE & 0.219 & 0.171 & 0.157 & 0.155 & 0.152 & 0.139 & 0.142 \\
   							& COV & 0.877 & 0.851 & 0.799 & 0.749 & 0.694 & 0.621 & 0.517 \\ \hline
   							
     $\hat{\mu}_{\mathrm{MLE}}$ & Bias & -0.129 & -0.064 & -0.030 & -0.026 & -0.024 & -0.012 & -0.021 \\
     						& RMSE & 0.489 & 0.291 & 0.288 & 0.222 & 0.178 & 0.152 & 0.128\\ 
   							& COV & 0.780 & 0.830 & 0.856 & 0.884 & 0.884 & 0.904 & 0.901 \\ \hline
   							
   	 $\hat{\mu}_{\mathrm{DS-P-BR}}$ & Bias & -0.132 & -0.091  & -0.074 & -0.065 & -0.057 & -0.041 & -0.040 \\ 
     						& RMSE       & 0.262 & 0.182  & 0.151 & 0.131  & 0.116 & 0.092 & 0.084\\ 
   							& COV   & 0.749 & 0.779  & 0.788 & 0.803  & 0.814 & 0.841 & 0.825 \\ \hline
   							
   	   	 $\hat{\mu}_{\mathrm{DS-LASSO}}$ & Bias   & -0.007 & -0.008  & -0.004 & -0.008 & -0.015 & -0.007 &  -0.015\\ 
     						& RMSE       		& 0.323 & 0.226  & 0.217 & 0.182  & 0.155 & 0.135 & 0.120 \\ 
   							& COV   		& 0.890 & 0.919  & 0.911 & 0.920  & 0.907 & 0.926 &  0.909\\ \hline	\hline

   \end{tabular}
    }

   \label{TableAsymptotic}
   \end{table}
}


  \scriptsize{ \textnormal{NOTE: OR: Outcome Regression, PS: Propensity Score. } }
  
  \normalsize

 \newpage

   {\renewcommand{\arraystretch}{1.3}
    \begin{table}[ht!]
   \centering
  \captionsetup{justification=centering}
   \caption{The effect of life expectancy on GDP growth: estimates of the ATE, their asymptotic standard error estimates (ASSE) and $95\%$ confidence intervals (CI).}
   \label{DataAnalysisTable}
    \resizebox{9cm}{!}{ 
      \begin{tabular}{l  r r l  }
   \hline\hline
   Estimator & ATE  & ASSE & CI   \\  \hline
   \textbf{$p=9$}      &      &        &    \\ 
$\hat{\mu}_\mathrm{OR}(\hat{\beta}_\mathrm{OLS})$	&	-5.386 & 0.837 & $[-7.02;-3.74]$	\\ 
$\hat{\mu}_\mathrm{Pop-IPTW}(\hat{\gamma}_\mathrm{MLE})$	&	1.678 &  0.423 & $[0.84; 2.50]$	\\ 
$\hat{\mu}_\mathrm{OR}(\hat{\beta}_\mathrm{LASSO})$  	&	 -2.228 &  0.512 & $[-3.23 ;  -1.22]$	\\ 
$\hat{\mu}_\mathrm{Pop-IPTW}(\hat{\gamma}_\mathrm{LASSO})$ 	&	1.373 & 0.475 & $[0.44;  2.30]$ 	\\ 
$\hat{\mu}_\mathrm{MLE}$	&	-5.391 & 0.879 & $[-7.11;  -3.66 ]$ 	\\
 $\hat{\mu}_\mathrm{LASSO}$	&	-2.406 & 0.622 & $[ -3.62 ;  -1.18 ]$		\\ 
$\hat{\mu}_\mathrm{DS-LASSO}$	&	-5.149 &  0.852 & $[-6.82; -3.47]$		\\
$\hat{\mu}_\mathrm{Post-LASSO}$ 	&	-5.174 & 0.858 & $[-6.85;  -3.49]$\\ 
$\hat{\mu}_\mathrm{P-BR}$	&	 -2.003 &  0.492 & $[-2.96;  -1.03]$ 	\\ 
 $\hat{\mu}_\mathrm{DS-P-BR}$	&	-3.578 & 0.578 & $[-4.71;  -2.44]$ 	\\
 
   \hline 

     \textbf{$p=63$}      &      &        &    \\ 
$\hat{\mu}_\mathrm{OR}(\hat{\beta}_\mathrm{OLS})$	&	 812.6 &   230.2 &  $[361.3;  1263.8]$ \\ 
$\hat{\mu}_\mathrm{Pop-IPTW}(\hat{\gamma}_\mathrm{MLE})$	& 1.721 &   0.421 & $[0.89; 2.54]$ 	\\ 
$\hat{\mu}_\mathrm{OR}(\hat{\beta}_\mathrm{LASSO})$  	&	 -6.013 &   1.275 & $[-8.51; -3.51 ]$	\\ 
$\hat{\mu}_\mathrm{Pop-IPTW}(\hat{\gamma}_\mathrm{LASSO})$ 	&	1.274 &  0.490 & $[0.31;2.23]$\\ 
$\hat{\mu}_\mathrm{MLE}$	&	812.6 &   230.2 & $[361.3; 1263.8]$	\\
 $\hat{\mu}_\mathrm{LASSO}$	&	-6.188 & 1.314 & $[-8.76; -3.61]$	\\ 
$\hat{\mu}_\mathrm{DS-LASSO}$	&	-13.27 &  2.089 & $[-17.36; -9.17]$ 	\\
$\hat{\mu}_\mathrm{Post-LASSO}$ 	&	-12.89 & 2.053 & $[-16.92; -8.87]$   \\ 
$\hat{\mu}_\mathrm{P-BR}$	&	-1.813 & 0.562 & $[-2.91; -0.71 ]$ \\ 
 $\hat{\mu}_\mathrm{DS-P-BR}$	& -28.80 & 5.214 & $[-39.02;-18.58]$\\  \hline\hline
   \end{tabular}
   }
   \end{table}

\end{document}